\documentclass[aps,prl,superscriptaddress,showpacs,floatfix,notitlepage,twocolumn,preprintnumbers]{revtex4-2}
\usepackage{amsmath,graphicx,float,csquotes,mdframed,appendix}
\usepackage{array}
\usepackage{booktabs}
\usepackage[colorlinks=true,linkcolor=blue,citecolor=blue, urlcolor=blue]{hyperref}
\usepackage[normalem]{ulem}
\usepackage{multirow,booktabs}
\usepackage{overpic}

\newcommand{\traj}{\emph{Trajectum}}

\begin{document}
 
\title{Yoctosecond imaging of the ground state of $^{129}$Xe at the Large Hadron Collider}

\author{Giuliano Giacalone}
\affiliation{Theoretical Physics Department, CERN, CH-1211 Genève 23, Switzerland}

\author{Govert Nijs}
\affiliation{Theoretical Physics Department, CERN, CH-1211 Genève 23, Switzerland}

\author{Wilke van der Schee}
\affiliation{Theoretical Physics Department, CERN, CH-1211 Genève 23, Switzerland}
\affiliation{Institute for Theoretical Physics, Utrecht University, 3584 CC Utrecht, The Netherlands}
\affiliation{NIKHEF, Amsterdam, The Netherlands}

\begin{abstract}
Imaging a quantum many-body system requires probes that resolve the coordinates of its constituents in sufficiently large event samples, allowing measurements of correlation functions \cite{Serwane:2011gjh,deJongh:2024pmo,Xiang:2024isi,Yao:2024rew}. High-energy nuclear collisions provide this opportunity on the nuclear scale \cite{STAR:2024wgy}, enabling features of colliding ions, such as their deformation, to be probed through particle correlation observables \cite{Jia:2022ozr,Giacalone:2023hwk}. However, a quantitative extraction of the correlation properties of nuclei from these measurements is still lacking. Here we show that this is possible for the nucleus $^{129}$Xe using Bayesian inference methods. We combine a deformed-rotor description of the colliding nuclei, which encodes the many-body dynamics of constituent neutrons and protons, with hydrodynamic simulations of the ensuing collision evolution. From a combined global analysis of Large Hadron Collider data on Xe-Xe and Pb-Pb collisions, we then infer that the shape of $^{129}$Xe is nearly maximally triaxial, which aligns with mean-field results for xenon isotopes away from shell closure \cite{Bally:2021qys,Grams:2026jwg}. From this we evaluate two- and three-particle correlations in the nuclear ground state to provide new constraints for \textit{ab initio} methods in nuclear theory. We establish thus collider experiments as a means of quantifying correlations of protons and neutrons arising from residual forces of quantum chromodynamics.
\end{abstract}

\preprint{CERN-TH-2026-124}

\maketitle

\paragraph{\textbf{Imaging quantum systems.}}

A single high-resolution photograph may be enough to image a static classical system. However, when quantum mechanics governs spatial resolution alone is not sufficient, because a quantum system cannot be decoded from a single shot. Snapshots of the many-body wave function become informative only when collected over a large ensemble, to exhibit correlations among constituents. Imaging in quantum mechanics is thus fundamentally tied to the measurement and interpretation of correlation functions. 

This notion has recently acquired prominence in the study of ultracold quantum gases, where advances in quantum gas microscopy have enabled the analysis of atomic ensembles with micrometer-scale single-atom resolution. This has allowed direct measurements of correlation functions, revealing phenomena such as the formation of Cooper pairs \cite{Holten:2021pex}, the signatures of quantum statistics \cite{Holten:2020jrt,deJongh:2024pmo,Xiang:2024isi}, or the formation of molecules driven by inter-atom interactions \cite{Brandstetter:2023jsy,Brandstetter:2024gur,Yao:2024rew}.

Going one billion times smaller leads us to the scale of atomic nuclei. Is it possible to perform the same type of imaging for a nucleus? Can we reconstruct correlations of constituent protons and neutrons (collectively denoted as \textit{nucleons}) by effectively taking photographs of their positions? A growing body of work \cite{Jia:2022ozr}, driven by recent observations made at colliders \cite{STAR:2015mki,ALICE:2018lao,STAR:2021mii,ATLAS:2022dov,STAR:2024wgy,ALICE:2024nqd,STAR:2025elk,ATLAS:2025nnt,ALICE:2025luc,CMS:2025tga,CMS:2025opi}, suggests that this is indeed possible through the study of multi-particle correlation observables \cite{Ollitrault:2023wjk} in the final states of ultrarelativistic ion–ion collisions at facilities such as the BNL Relativistic Heavy Ion Collider (RHIC) or the CERN Large Hadron Collider (LHC).

\begin{figure*}[t]
\centering
\includegraphics[width=1\linewidth]{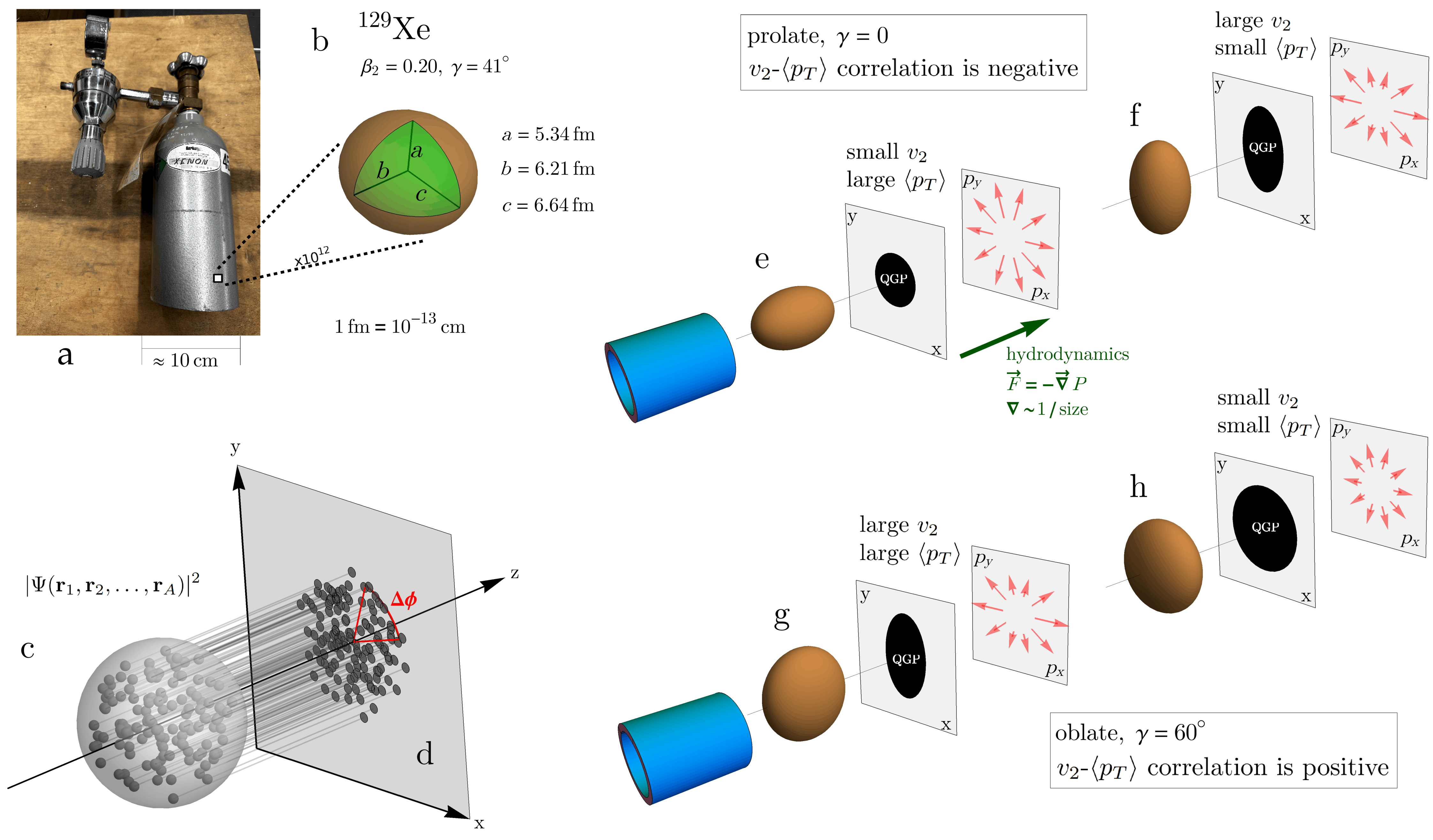}
\caption{\label{fig:1} \textbf{Reconstructed rotor shape for xenon-129 and the collider imaging technique.} \textbf{a} Photograph of the bottle of xenon gas used for injection in the LINAC3 facility of the accelerator complex at CERN. \textbf{b} Zooming to nuclear scales, these nuclei can be visualized as kiwi-shaped deformed spheroids having three unequal axes. The lengths of the axes are here determined from the surface parametrization of Eq.~(\ref{eq:Rthetaphi}) using parameters $\beta_2$ and $\gamma$ extracted in this work from LHC data. \textbf{c-d} In an ultrarelativistic collision, one probes the transverse projections of the nucleons coming from the initial $^{129}$Xe nucleus, whose spatial distribution shapes the subsequent quark-gluon plasma (QGP). In the classical rotor approximation, the deformed shape is projected onto the transverse plane, leading to characteristic effects on the QGP geometry in the limit of fully head-on collisions that we discuss here. For a prolate deformation with $\gamma=0$, the projected shape can vary from small and circular \textbf{e}, leading to reduced elliptic momentum anisotropy, $v_2 = (p_x^2-p_y^2 ) / (p_x^2+p_y^2)$, but higher mean transverse momentum, $\langle p_T \rangle = \sqrt{p_x^2+p_y^2}$, to large and elliptical \textbf{f}, leading to reduced $\langle p_T \rangle$ but larger $v_2$. For an oblate nucleus, 
the smallest area corresponding to a larger $\langle p_T \rangle$ is instead elliptical \textbf{g}, leading at the same time to significant $v_2$, while the maximal area is circular \textbf{h}, leading to both small $\langle p_T \rangle$ and small $v_2$. 
}
\end{figure*}

In that regime, the collision acts as a sudden probe of the nuclear state. The passage time between two nuclei is below the yoctosecond scale ($10^{-24}\,s$), so that nucleon coordinates are \textit{frozen} during the interaction: each collision samples a configuration drawn from the ground-state many-body wave function [Figs.~\ref{fig:1}c-d]. At the same time, the probe is inherently collective: in fully-overlapping (or ultra-central) events, all nucleons are probed simultaneously. Through hydrodynamic flow, final-state multi-particle correlations are then sensitive to the average arrangement and correlations among incoming nucleons \cite{Giacalone:2023hwk}.

In this work, we image the nucleus $^{129}$Xe using data from a dedicated run of Xe-Xe collisions at the LHC. We describe this nucleus within a semiclassical rotor picture featuring a quadrupole-deformed triaxial shape [see Fig.~\ref{fig:1}b], which effectively encodes long-range angular correlations among nucleons in the lab frame that we infer from experimental data through a global Bayesian analysis. From the resulting posterior constraints, we obtain measurements of two- and three-nucleon angular correlations in the ground state of $^{129}$Xe, and discuss the broader implications and potential of this result.

\paragraph{\textbf{Many-body correlations and rotor model.}}

The remarkable phenomenological success of perturbative quantum chromodynamics (QCD) computations is based on the notion of high-energy factorization. Here an elementary scattering process is separated into a non-perturbative part describing the slow, low-energy structure of the probe, conventionally encoded in parton distribution functions, and a short-distance scattering amplitude that is calculable in perturbation theory \cite{Gelis:2015gza}. In high-energy nuclear collisions, an analogous picture appears to arise in which parton distributions are replaced by the spatial configuration of the incoming nucleons \cite{Miller:2007ri}, while elementary interactions are effectively subsumed into the emergence of a hydrodynamic medium, the quark-gluon plasma (QGP) \cite{Busza:2018rrf}.

Therefore, the primordial input to a high-energy nuclear collision comes from low-energy nuclear structure. Denoting by $|\Psi\rangle$ the nuclear ground state, one needs its associated many-body probability density
\begin{equation}
|\Psi(\mathbf{r}_1,\ldots,\mathbf{r}_A)|^2 = \big|\langle \mathbf{r}_1, \ldots , \mathbf{r}_A | \Psi \rangle\big|^2 .
\end{equation}
The relevant information from such a system is encoded in (spin- and isospin-averaged) many-body densities,
\begin{equation}
\label{eq:rho(n)}
\rho^{(n)}(\mathbf{r}_1,\ldots,\mathbf{r}_n) = \int d\mathbf{r}_{n+1}\cdots d\mathbf{r}_A \, |\Psi(\mathbf{r}_1,\ldots,\mathbf{r}_A)|^2 .
\end{equation}
Atomic nuclei are intrinsically strongly correlated systems and their properties can change abruptly even between neighboring isotopes. It is now established that, by probing event-by-event snapshots of the underlying many-body configurations, high-energy collisions are sensitive to detailed two-, three-, and possibly higher-body nucleon correlations \cite{Giacalone:2023hwk}. It is therefore of fundamental interest to ask whether these processes can be used as quantitative probes of these correlated quantum many-body states. This is our driving question.

To model a correlated density of nucleons in a way that can be fitted to the data, we borrow intuition from mean-field calculations of nuclear structure \cite{Bender:2003jk}. We construct the many-body probability density, $|\Psi|^2$, as the orientation average of a deformed density of nucleons, denoted $\rho_\Omega({\bf r})$, which in each realization of the nucleus has a random orientation with respect to the lab frame and from which nucleons are sampled independently. Operationally, in each collision we sample \textit{(i)} a set of three Euler angles, $\Omega$, that define the orientation of the density $\rho_\Omega$ in the laboratory frame, and \textit{(ii)} a set of $A$ independent nucleon positions $\{ {\bf r}_1\ldots{\bf r}_A\}$ from that rotated distribution. 
All correlations among nucleons emerge then as a consequence of the average over orientations, which is performed once observables in the lab frame are evaluated. This approach approximates the quantum projection technique used to project intrinsic deformed states onto good quantum numbers of angular momentum in the mean-field frameworks \cite{Dobaczewski:2025rdi,Ke:2025tyv,Bofos:2026nmg}.

More specifically, for the intrinsic density of $^{129}$Xe and $^{208}$Pb we consider a realistic Woods–Saxon profile,
\begin{equation}
\label{eq:rhoWS}
\rho^{\mathrm{WS}}(\mathbf{r}) \propto \left [ 1 + \exp\left ( \frac{r-R(\theta,\phi)}{a} \right ) \right ]^{-1} ,
\end{equation}
where $a$ is the diffuseness of the nuclear surface, which we deform via a standard real-form spherical-harmonic expansion with Bohr deformation parameters,
\begin{align}
\label{eq:Rthetaphi}
\nonumber & R(\theta,\phi)= \\ 
&R_0\biggl (1+\beta_2 \biggl [\cos(\gamma) Y_2^0(\theta) +\sin(\gamma) Y_2^2(\theta,\phi) \biggr] +\beta_3Y_{3}^0(\theta) \biggr)  .
\end{align}
Here, $R_0$ gives the half-width radius, $\beta_2$ gives the magnitude of the quadrupole deformation of the density, $\gamma \in [0,60^\circ]$ determines whether the shape is prolate ($\gamma=0$), oblate ($\gamma=60^\circ$), or triaxial (three unequal axes), and $\beta_3$ is the magnitude of the octupole deformation, leading to a reflection-asymmetric shape. An example of a triaxial density for $\beta_3=0$ is in Fig.~\ref{fig:1}b. We denote now by $\rho^{\mathrm{WS}}_\Omega(\mathbf{r})$ the density of nucleons rotated by Euler angles $\Omega$ in the laboratory frame. Neglecting effects of non-zero angular momentum in the ground states, averaging over orientations restores spherical symmetry. The resulting one-body density is thus
\begin{equation}
\rho^{(1)}(\mathbf{r}_1) = \frac{1}{8\pi^2} \int d\Omega \,\rho^{\mathrm{WS}}_\Omega(\mathbf{r}_1) ,
\end{equation}
which depends only on $|{\bf r}_1|$. At the two-body level, when $\beta_2\neq0$ the collective alignment of nucleons in the intrinsic frame generates a non-factorizable correlation in the lab,
\begin{align}
\nonumber \rho^{(2)}(\mathbf{r}_1,\mathbf{r}_2) =  \frac{1}{8\pi^2}\int d\Omega \,\rho^{\mathrm{WS}}_\Omega&(\mathbf{r}_1)\,\rho^{\mathrm{WS}}_\Omega(\mathbf{r}_2) \, \\
&\neq  \rho^{(1)}({\bf r}_1)\rho^{(1)}({\bf r}_2) . 
\end{align}
Higher-order correlation functions follow analogously.

The rotor model should then be understood as a means of introducing a spectrum of Fourier harmonics in the two- and higher-body distributions of nucleons \cite{Duguet:2025hwi,Blaizot:2025scr}. Denoting ${\bf r}=({\bf r}_\perp, z)$ and ${\bf r}_\perp=(r_\perp, \phi)$, the orientation average of the rotor gives a characteristic contribution
\begin{align}
\nonumber    \int dz_1 \,dz_2 ~ \bigl [\rho^{(2)}({\bf r}_1, {\bf r}_2)&-\rho^{(1)}({\bf r_1})\rho^{(1)}({\bf r}_2 ) \bigr ] \\
   &\propto \sum_n \beta_n^2 \,r_{1\perp}^n \, r_{2\perp}^n  \cos(n\,\Delta\phi),
\end{align}
where $n$ runs over the Fourier components that are included in the expansion of Eq.~(\ref{eq:Rthetaphi}), and $\Delta \phi= \phi_1-\phi_2$, also illustrated in Fig.~\ref{fig:1}d. These harmonic modulations impact the multi-particle correlations measured in the collision final states, opening a bridge between collider observables and many-body nuclear dynamics. We now discuss how this is enabled within the hydrodynamic description of heavy-ion collisions.
 
\paragraph{\textbf{Hydrodynamic response as an imaging tool.}}
On an event-by-event basis, the QGP undergoes a collective hydrodynamic expansion before fragmenting into final-state hadrons. The notion of an initial-state geometry is key in this context. Here, we focus on central collisions at small impact parameters, where the colliding nuclei overlap nearly completely. The characteristic resolution scale of gluon-mediated processes responsible for transverse energy deposition, $\mathcal{O}$(1GeV)$\sim$$\mathcal{O}$(0.1fm), is far smaller than the geometric scale set by the nuclear radius, $\mathcal{O}$(5fm). Because of this strong scale separation, partonic interactions do not themselves alter the long-range geometry of the interaction region. Rather, the large-scale structure of the QGP is determined by the transverse positions of the incoming nucleons that source these partons, and hence by the low-energy nuclear properties that govern those positions.

This becomes especially striking within the picture of the classical rotor. We follow Fig.~\ref{fig:1}e-h. The non-spherical shape of the colliding nuclei is imprinted directly into the transverse geometry of the QGP. Once formed, the plasma expands hydrodynamically and its evolution is governed by local conservation of momentum. Specifically, the Euler equations imply that pressure-gradient forces drive fluid acceleration in directions determined by their spatial distribution. This has two major consequences, illustrated in the figure.

First, the deformed shape impacts the anisotropic flow of matter in the transverse plane \cite{Ollitrault:2023wjk}. We measure a spectrum of hadrons in the midrapidity detector slice as a function of the azimuthal angle, $dN_{\rm hadrons}/d^2{\bf p}_T$, where ${\bf p}_T = (p_x, p_y)=(p_T,\phi_p)$. The anisotropic flow is defined by the Fourier coefficients,
\begin{equation}
\label{eq:Vn}
V_n  \propto \int d^2{\bf p}_T\,e^{i n \phi_p} \, \frac{dN_{\rm hadrons}}{d^2{\bf p}_T}, \hspace{20pt} |V_n|=v_n.
\end{equation}
Quadrupole deformation ($\beta_2$) enhances the elliptical anisotropy of the QGP, leading to enhanced elliptic flow, $v_2$, while octupole deformation ($\beta_3$) enhances triangular flow, $v_3$, and so on \cite{Jia:2021tzt}. These $v_n$ values, accessible experimentally, are then sensitive to the rotor shape.

\begin{figure*}[t]
\centering
\includegraphics[width=\textwidth]{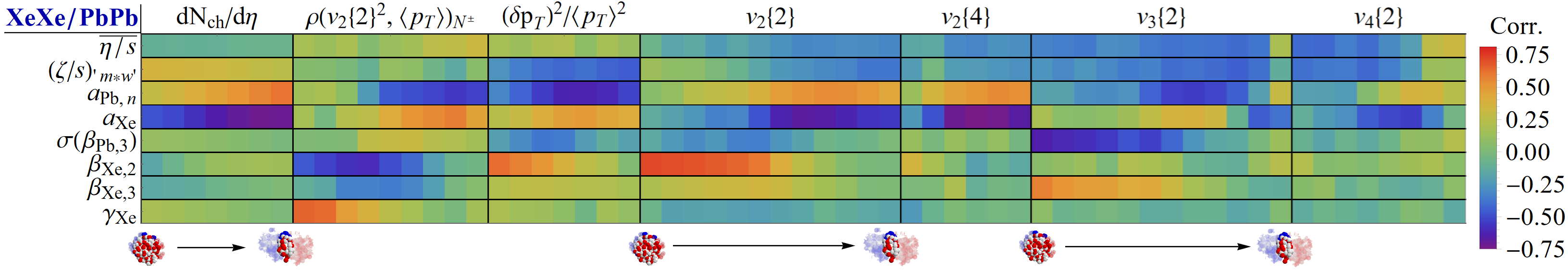}
\caption{\label{fig:2} \textbf{Sensitivity matrix of observables computed in Xe-Xe collisions relative to Pb-Pb collisions.} The final-state observables, which are listed in the top of the table, are ratios of quantities taken between Xe-Xe and Pb-Pb systems, with the exception of the $\rho(v_2\{2\}^2,\langle p_T \rangle)$ correlator for which we take a difference (to avoid divisions by zero). For each observable, we plot the corresponding correlation with a given model parameter, where we select here particularly representative parameters. For each observable and parameter, going from left to right means increasing the centrality class in which the correlation is computed. Results for absolute observables in Xe-Xe collisions and for all model parameters are provided in the supplement.
}
\end{figure*}

Secondly, the anisotropy of the QGP is coupled non-trivially to its overall size. Through the Euler relation, the transverse size of the medium is reflected in the strength of radial flow \cite{Bozek:2012fw}, which can be quantified by the mean transverse momentum of produced hadrons,
\begin{equation}
\label{eq:[pT]}
\langle p_T \rangle \propto  \int d^2{\bf p}_T\,p_T \, \frac{dN_{\rm hadrons}}{d^2{\bf p}_T}.
\end{equation}
A key feature is that $\langle p_T \rangle$ scales inversely with the transverse size of the QGP: larger systems produce weaker gradients and radial flow, and hence smaller $\langle p_T \rangle$, while smaller systems produce stronger radial flow and larger $\langle p_T \rangle$. For prolate nuclei in Figs.~\ref{fig:1}e-f ($\beta_2 > 0$, $\gamma = 0$), the maximum overlap area occurs when the symmetry axis is aligned perpendicular to the beam direction ($z$). In this configuration, the transverse area is largest but also most elliptic, leading to a characteristic anti-correlation between $v_2$ and $\langle p_T \rangle$ \cite{Giacalone:2019pca}. In contrast, for oblate nuclei in Figs.~\ref{fig:1}g-h ($\beta_2 > 0$, $\gamma = 60^\circ$), the maximal overlap area is nearly circular. In this case, the overlap area and ellipticity vary in concert, instead leading to a positive correlation between $v_2$ and $\langle p_T \rangle$ \cite{Bally:2021qys,Jia:2021qyu}.

Therefore, in the intrinsic rotor picture, the impact of deformations on the final-state observables can be assessed by studying the interplay between the anisotropic flow coefficients, $v_n$, and the mean transverse momentum, $\langle p_T \rangle$. This combined analysis provides a powerful experimental handle for inferring quantitative information about nuclei from collider measurements. 

\begin{figure*}[t]
\centering
\includegraphics[width=\textwidth]{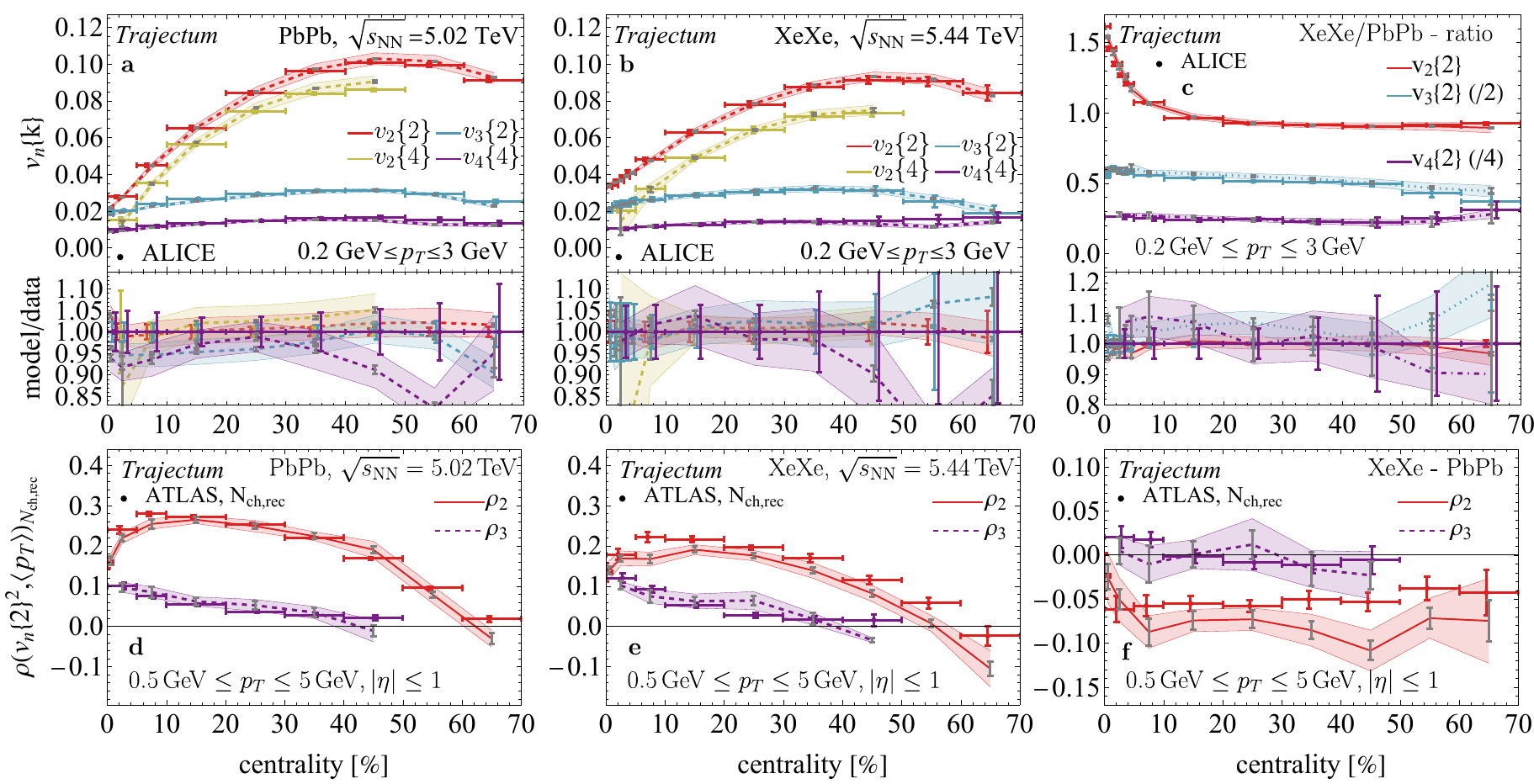}
\caption{\label{fig:3} \textbf{Model evaluations resulting from a global Bayesian analysis of LHC Pb-Pb and Xe-Xe data.} We present results for selected observables carrying a strong sensitivity to nuclear shape. The first row of panels displays anisotropic flow coefficients in Pb-Pb collisions \textbf{a} and in Xe-Xe collisions \textbf{b}. Ratios of such observables are shown in panel \textbf{c}. The second row shows results for $v_n^2$-$\langle p_T \rangle$ correlations, again in Pb-Pb collisions \textbf{d}, Xe-Xe collisions \textbf{e}, as well as their difference \textbf{f}. For the Trajectum results we show both statistical (bars) and systematical (bands) theoretical uncertainties. 
}
\end{figure*}

\paragraph{\textbf{Bayesian analysis of LHC data.}}

We perform the first combined Bayesian fit of LHC Xe-Xe and Pb-Pb data within the \textit{Trajectum} framework of high-energy nuclear collisions \cite{Nijs:2020ors,Nijs:2020roc}. We focus on quantities such as those in Eqs.~(\ref{eq:Vn}) and (\ref{eq:[pT]}) which are not differential in $p_T$ and are more sensitive to initial-state effects. When needed, we inflate the experimental uncertainties to fit data points whose relative uncertainties are below 4\%, which sets essentially our target precision for the fits. The observables are fitted as a function of the collision centrality, whereby 0-1\% centrality corresponds to ultra-central events. We fit yields, average transverse momenta, and multi-particle correlation observables. Among the latter, information about the quadrupole deformation of $^{129}$Xe is especially carried by the root mean squared anisotropic flow coefficients, $v_n\{2\}\equiv \sqrt{\langle v_n^2 \rangle}$, and the linear correlation between the squared elliptic flow and the mean transverse momentum, which we denote by $\rho(v_2\{2\}^2,\langle p_T \rangle)$ \cite{Bozek:2016yoj}. To maximize the sensitivity to nuclear structure effects, we look at relative observables. For example, while the value of $v_2\{2\}$ in a single collision system such as Xe-Xe collision depends substantially on QGP medium properties, such as transport coefficients, once we take a ratio of the Xe-Xe data point with the Pb-Pb data point, we eliminate most of such dependencies and neatly isolate initial-geometry effects. The lead nucleus is an excellent baseline for this purpose, as it has essentially no quadrupole deformation.

Our goal is to demonstrate that a rigorous determination of $\gamma$ is possible. As argued and numerically validated in several studies, $v_2\{2\}^2$ has a leading sensitivity to $\beta_2^2$, while $\rho(v_2\{2\}^2,\langle p_T \rangle)$ to the combination $\beta_2^3 \cos(3\gamma)$ \cite{Jia:2021qyu,Mehrabpour:2026yuc}. Therefore, one can first use the elliptic flow to infer $\beta_2$, and with this knowledge proceed to extract $\gamma$ from the $v_2^2$-$\langle p_T \rangle$ correlator. However, a Bayesian analysis such as ours has not yet established the level of theoretical uncertainty with which $\beta_2$ and $\gamma$ can be extracted. This is our main objective. 

In Fig.~\ref{fig:2}, we display a matrix of sensitivities of observables to model parameters for different centrality classes. The quantities shown are relative observables between the Xe-Xe and Pb-Pb systems. Our first remark is that we validate the expectation that the effects of QGP-medium properties are only marginal. In the first two rows, there is only a weak correlation between observables and the magnitude of the shear viscosity, $\bar \eta/s$, or of the bulk viscosity, $\zeta/s$, of the QGP, without any clear centrality dependence. This changes dramatically in subsequent rows of the plot, showcasing correlations with nuclear structure parameters. The diffuseness parameters are strongly correlated with the centrality, as expected, and there is a strong anti-correlation between $a_{\rm Xe}$ and $a_{\rm Pb}$ due to the ratio. Concerning the shape parameters, the strongest sensitivities are instead in central collisions, also as expected. This also implies that it should be possible to simultaneously determine both $a$ and $\beta_2$ from the global fits. Getting to our main target, we confirm that $\rho(v_2\{2\}^2,\langle p_T \rangle)$ is the only observable with a leading sensitivity to $\gamma$ in the most central collisions.

Proceeding to the global fit of the experimental data, our key results are provided in Fig.~\ref{fig:3}. We stress that the fit is excellent, with all theoretical curves lying essentially within $\pm5\%$ of the experimental values across a broad centrality range. The data points that are most sensitive to nuclear structure effects, as anticipated by the sensitivity matrix of Fig.~\ref{fig:2}, are also captured very accurately. This suggests that nuclear structure information can be extracted from data, as we now discuss.

\paragraph{\textbf{Two- and three-body correlations in $^{129}$Xe}.}

\begin{figure}[t]
\centering
\includegraphics[width=\columnwidth]{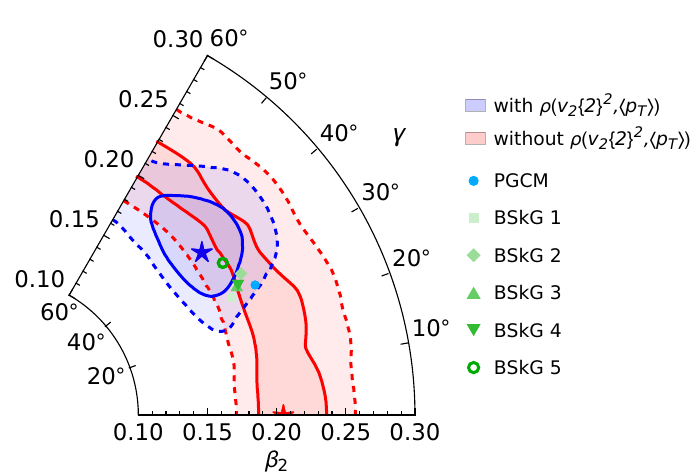}
\caption{\label{fig:4}
\textbf{Posterior distribution of $\beta_2$ and $\gamma$ parameters for $^{129}$Xe.} We show with (blue) and without (red) the inclusion of $\rho(v_2\{2\}^2,\langle p_T \rangle)$ in the fits. Stars indicate Maximum A Posteriori values (see also Tab.~\ref{tab:params} in the supplement). The other symbols correspond to deformations obtained from intrinsic one-body densities within Projected Generator Coordinate Method (PGCM \cite{Bally:2021qys}) and Brussels-Skyrme-on-a-Grid 1-5 (BSkG1-5 \cite{Grams:2026jwg}) mean-field calculations.
}
\end{figure}

The inferred posterior distributions of the deformation parameters are given in Fig.~\ref{fig:4}. We find a well-deformed shape ($\beta_2\approx0.20$), with a slight preference for oblateness ($\gamma\approx40^\circ$). As shown in the figure, this result aligns well with mean-field calculations for the ground state of $^{129}$Xe. As argued, the shape provides an effective tool for modeling nontrivial spatial correlations of nucleons. We now translate the information encoded by the shapes into well-defined observables that can be connected to \textit{ab initio} approaches to the nuclear many-body problem.

To this end we introduce the one-body transverse radius and eccentricity operators, respectively, defined by
\begin{equation}
   R_\ell({\bf r}) =r_\perp^\ell, \hspace{20pt} \mathcal{E}_\ell ({\bf r})= r_\perp^{|\ell|} e^{i\ell\phi} \, .
\end{equation}
For a given $n$-body operator, $\mathcal{O}({\bf r}_1 \ldots {\bf r}_n)$, we introduce the following notation for its expectation
\begin{equation}
\label{eq:Oavg}
\langle \mathcal{O} \rangle_n = \int d\mathbf{r}_{1}\cdots d\mathbf{r}_n \, \rho^{(n)}({\bf r}_1\ldots{\bf r}_n) \mathcal{O}({\bf r}_1 \ldots {\bf r}_n),
\end{equation}
where the many-body densities are defined in Eq.~(\ref{eq:rho(n)}). From the inferred shape parameters of $^{129}$Xe, we evaluate ground-state expectations of key many-body operators.

The first is the mean-squared nuclear eccentricity \cite{Duguet:2025hwi,Bofos:2026huw}, which quantifies the second harmonic modulation of the ground-state two-body distribution \cite{Blaizot:2025bfu}
\begin{equation}
\label{eq:E2E-2}
    \frac{\langle\mathcal{E}_2 \,\mathcal{E}_{-2} \rangle_2}{\langle R_2 \rangle_1^2} \equiv \frac{\langle r_{1\perp}^2 r_{2\perp}^2 e^{i2(\phi_1-\phi_2)} \rangle_2}{\langle r_{1\perp}^2\rangle_1^2} = 0.0131(1)_\text{stat}(19)_\text{syst}\,.
\end{equation}
The expectation value of this two-body operator is the leading contribution to $v_2\{2\}^2$ coming from two-body nuclear correlations, and in the rotor picture is proportional to $\beta_2^2$ \cite{Blaizot:2025scr,Duguet:2025hwi}. The second is the covariance of the mean squared anisotropy and the nuclear radius, quantifying how the shape and the size of the nucleus are correlated. It involves a three-body operator expectation
\begin{equation}
\frac{\langle R_2 \,\mathcal{E}_2\,\mathcal{E}_{-2} \rangle_3}{\langle R_2\rangle_1^3} - \frac{\langle\mathcal{E}_2 \, \mathcal{E}_{-2} \rangle_2}{\langle R_2 \rangle_1^2} = 0.01(1)_\text{stat}(30)_\text{syst} \times 10^{-3} \,.
\end{equation}
In the rotor picture, and to leading order in $\beta_2$, this quantity is directly proportional to $\beta_2^3 \cos(3\gamma)$ \cite{Mehrabpour:2026yuc}, hence its closeness to zero when the rotor is triaxial ($\gamma\approx30^\circ$).

These are the first experimental determinations of these quantities. In the supplement, we perform different iterations of the global Bayesian inference, where we reduce the size of experimental errors down to 1\% relative errors, we add the fit of the $R_0$ parameter of the Woods-Saxon density, and where we study the impact of different centrality definitions used in the experiments. Our conclusion is that the extraction of $\beta_2$ and $\gamma$ is robust, and that the resulting theoretical uncertainties should be taken seriously and can be used to constrain nuclear structure theory. In contrast, we find that the determination of the nuclear skin diffuseness parameters is less robust. This suggests that inferring the radius of a nucleus is more difficult than obtaining many-body properties tied to angular correlations of nucleons.

The present model assumes that all correlations within the nucleus can be efficiently captured through the deformed rotor picture.  In the future, these extractions can therefore be improved by generalizing such a model with the inclusion of pairing, Pauli exclusion, and short-range repulsion effects, which leave distinct imprints on the many-body nucleon distributions \cite{Blaizot:2025bfu,Sun:2026yrr}.

Before concluding, we note that, including in the fit fluctuating $\beta_2$ and $\beta_3$ parameters for $^{208}$Pb with a Gaussian distribution (with zero average), we obtain a robust determination of a standard deviation $\sigma(\beta_{\rm Pb,3}) = 0.127\pm 0.023$ (90\% confidence) for the octupole deformation, while quadrupole correlations are consistent with zero. A significant octupole correlation in $^{208}$Pb aligns with a recent analysis done in the context of Coulomb excitation measurements for this nucleus \cite{Henderson:2025scq}. We also find a significant (average) $\beta_3$ parameter for $^{129}$Xe ($\beta_{{\rm Xe},3}=0.09\pm 0.05$), a rather intriguing finding with potential connections to nuclear Schiff moments \cite{Belley:2026bde}.

\paragraph{\textbf{Discussion.}}
Recent years have brought dramatic progress in understanding how the properties of colliding ions are encoded in final-state observables. However, a robust Bayesian framework for extracting nuclear-structure information from these measurements, with reliable uncertainty quantification, has been lacking. Here we have filled this gap, demonstrating ultrarelativistic collisions as a quantitative tool to study correlated nuclear ground-state structure. We have quantified two- and three-body angular correlations of nucleons in the ground state of a xenon isotope, encoded via a rotor description. This has a number of implications.

First, these observables provide new benchmarks for \textit{ab initio} calculations, triggering a broader program in which Bayesian constraints from collider data are coupled to Bayesian sensitivity studies of the low-energy constants of chiral effective field theory interactions \cite{Sun:2024iht}. Such a program would make it possible to evaluate the complementarity of low- and high-energy probes in constraining effective theories of low-energy QCD. Furthermore, three-body correlations and couplings may play a central role for dense matter \cite{Somasundaram:2024ykk}, such that this line of development may ultimately have implications for the physics of neutron stars and their mergers.

Further opportunities relate to new physics searches with atomic nuclei. A particularly compelling example is the determination of nuclear matrix elements (NME) for neutrinoless double-beta decay, which remain a bottleneck in both the interpretation of existing searches and the design of future experiments \cite{Agostini:2022zub}. These matrix elements are highly sensitive to many-body correlations, especially in the parent nuclei. Direct constraints on such nuclei from collider measurements could therefore help reduce the theoretical uncertainties on NME determinations \cite{Li:2025vdp}. Additionally, direct dark matter and neutrino searches involve neutron form factors that are probed in high-energy collisions \cite{AristizabalSierra:2019zmy}, while correlation effects related to nuclear deformations and accessible in high-energy smashups contribute to precision spectroscopy with nuclear clocks \cite{Caputo:2024doz}. Therefore, our analysis opens the way to a broader program in which collider experiments serve as precision instruments for imaging strongly correlated nuclear systems.

\section{Acknowledgment}

We thank Benjamin Bally, Thomas Duguet, Jiangyong Jia, Matt Luzum, Maxim Virta, Chunjian Zhang, You Zhou for useful discussions. We thank Reyes Alemany Fernandez and Detlef Kuchler for sharing the photograph of the xenon gas bottle. We thank Pepijn Demol and Wouter Ryssens for providing us with the BSkG1-5 results. The authors acknowledge the Kavli Institute for Theoretical Physics (KITP, Santa Barbara) and the Yukawa Institute for Theoretical Physics (YITP, Kyoto) for hospitality during the finalization of this work.

\bibliographystyle{apsrev4-1}
\bibliography{main,manual}

\section{methods}

\paragraph{\textbf{Trajectum framework and Bayesian analysis.}}

The \emph{Trajectum} framework used in this study contains four successive components: an initial-state construction, followed by pre-equilibrium dynamics, a hydrodynamic evolution stage, and a conversion to hadrons followed by hadronic transport with SMASH~\cite{Weil:2016zrk,dmytro_oliinychenko_2020_3742965}%
\@. Here, we briefly summarize the parameters entering the calculation, the observables used to constrain them, and the statistical procedure employed. More extensive technical details are given in Ref.~\cite{Nijs:2020roc}\@.

In the initial stage, nucleons are sampled inside the colliding nuclei according to Eq.~(\ref{eq:rhoWS}) of the main text, where for $^{208}$Pb we use separate parameters for protons and neutrons, and where for $^{129}$Xe we use a single description for both. 
In addition, the surface of the density is deformed via
\begin{equation}
R(\theta)=R\bigl(1+\beta_2\cos\gamma Y^0_2(\theta)+\beta_2\sin\gamma Y^2_2(\theta)+\beta_3Y^0_3(\theta)\bigr),
\end{equation}
where $R$ is the half-width radius, $\beta_2$ measures the quadrupole deformation, $\gamma$ measures the triaxiality, and $\beta_3$ measures the octupole deformation. For $^{208}$Pb, $\gamma = 0$ and we let $\beta_2$ and $\beta_3$ fluctuate around zero according to a Gaussian distribution with width
\begin{equation}
\sigma(\beta_n) \equiv \sqrt{\langle \beta_n^2\rangle}.
\end{equation}
In this work we do not enforce a short-range repulsion among nucleons (often done by a minimum nucleon-nucleon separation), mostly since the current implementation does not support such a minimum distance for triaxial nuclei.
The colliding nucleons (referred to as participants) are selected using the measured inelastic cross section $\sigma_{\rm NN}$, following Ref.~\cite{Moreland:2014oya}. %

Each participant is represented as $n_c$ constituents, each carrying a transverse Gaussian profile of width
\begin{equation}
v = 0.2~\mathrm{fm} + \chi_{\mathrm{struct}}\,(w-0.2~\mathrm{fm}).
\end{equation}
The constituent centers are themselves Gaussian distributed, so that the average nucleon profile has width $w$\@. Summing over participants and their constituents yields the thickness functions $\mathcal{T}_{A/B}$\@. The normalization of each constituent fluctuates as $N\gamma/n_c$, where $\gamma$ is drawn from a gamma distribution with mean unity and width $\sigma_{\mathrm{fluct}}\sqrt{n_c}$, where $N$ and $\sigma_\text{fluct}$ are parameters. The initial energy density $e$ is then obtained from the thickness functions by the T\raisebox{-0.5ex}{R}ENTo formula
\[
\tau e|_{\tau\rightarrow0^+} \propto \left(\frac{\mathcal{T}_A^p + \mathcal{T}_B^p}{2}\right)^{q/p},
\]
with parameters $p$ and $q$.

During a time interval $\tau_{\mathrm{hyd}}$, this initial energy density is evolved in two ways: by free streaming, and by a holographically motivated prescription~\cite{Nijs:2023yab}. These results are interpolated between these two limits using a continuous parameter $r_{\mathrm{hyd}}$, with $r_{\mathrm{hyd}}=0$ corresponding to free streaming and $r_{\mathrm{hyd}}=1$ to the holographic choice. This interpolation is then used to initialize the hydrodynamic simulation.

The stress-energy tensor is then propagated with second-order viscous hydrodynamics. Nine parameters control the transport sector:
$ (\eta/s)_{\mathrm{ave}}$,
$ (\eta/s)_{\mathrm{slope}}$,
$ (\eta/s)_{\delta\mathrm{slope}}$,
$ (\eta/s)_{0.8\,\mathrm{GeV}}$,
$ (\zeta/s)_{\max}$,
$ (\zeta/s)_{m\times w}$,
$ (\zeta/s)_{T_0}$,
$ \tau_\pi sT/\eta$,
and $\tau_{\pi\pi}/\tau_\pi$.
The first four specify the temperature dependence of the shear-viscosity-to-entropy ratio, namely its mean value, its slope between $150$ and $300$ MeV, the change in slope at higher temperature, and its value at $800$ MeV and above. The next three parameters determine the peak of the bulk viscosity, its effective width times height, and the temperature at which that peak is centered. The last two are dimensionless second-order combinations; in particular, $\tau_\pi sT/\eta$ controls the relaxation toward first-order viscous hydrodynamics. 
In this work we keep the equation-of-state coming from the HotQCD lattice computations fixed, as described in the original works by \cite{HotQCD:2014kol,Bernhard:2018hnz,Bernhard:2019bmu}.

At the switching temperature $T_{\mathrm{switch}}$, the fluid is converted to hadrons using the Cooper-Frye prescription with Pratt-Torrieri-Bernhard viscous corrections~\cite{Pratt:2010jt,Bernhard:2018hnz}\@. The resulting hadrons are subsequently evolved in SMASH, where all hadronic cross sections are multiplied by a factor $f_{\mathrm{SMASH}}$.

A final normalization parameter, $\text{cent}_\text{norm}$, adjusts the anchor point of the centrality (defining which multiplicity corresponds to 100\% centrality)\@. This is the only parameter assigned a non-flat prior; it is taken to follow a Gaussian prior with a width of 1\% around 100\%\@.

To focus on a robust determination of the shape of $^{129}$Xe this analysis uses a broad range of $p_T$-integrated observables for both $^{208}$Pb, $^{129}$Xe and importantly their ratio where possible. This includes the total hadronic cross sections in $^{208}$Pb+$^{208}$Pb collisions~\cite{ALICE:2022xir}, identified and unidentified yields $dN_{\mathrm{ch}}/dy$ and mean transverse momenta $\langle p_T\rangle$ for pions, kaons, protons and charged hadrons respectively \cite{ALICE:2015juo,ALICE:2019hno}. For xenon we similarly take the unidentified yields from \cite{ALICE:2018cpu}. Since this comes from the same experiment we model a mild cancellation of the uncertainty in the ratio by assuming the uncertainties are given by the maximum uncertainty instead of the quadratic sum. Integrated anisotropic flow coefficients for Pb-Pb and Xe-Xe collisions are given in respectively \cite{ALICE:2018rtz} and \cite{ALICE:2018lao}.

All these observables were from the ALICE collaboration, but subsequently we take the $\langle p_T \rangle$ fluctuations and the $\rho_2(v_2\{2\}^2,\langle p_T \rangle)$ correlator for both Xe-Xe and Pb-Pb from the ATLAS collaboration \cite{ATLAS:2022dov}.

The posterior distribution is then evaluated using Bayes' theorem,
\begin{equation}
P(x\mid y_{\mathrm{exp}})=
\frac{e^{-\Delta^2/2}}{\sqrt{(2\pi)^n\det\Sigma(x)}}\,P(x),
\label{eq:posterior}
\end{equation}
with a flat prior $P(x)$ for all parameters except $\text{cent}_\text{norm}$\@. The quadratic form entering the likelihood is
\begin{equation}
\Delta^2 = \bigl(y(x)-y_{\mathrm{exp}}\bigr)\cdot \Sigma(x)^{-1}\cdot \bigl(y(x)-y_{\mathrm{exp}}\bigr),
\label{eq:delta2}
\end{equation}
where $y(x)$ denotes the model prediction, $y_{\mathrm{exp}}$ the experimental data vector with $n$ entries, and $\Sigma(x)$ the sum of theoretical and experimental covariance matrices. Those covariance matrices are built as in Ref.~\cite{Bernhard:2018hnz}.

The standard workflow is to evaluate the model on a Latin-hypercube design and then train an emulator for $y(x)$, which is subsequently sampled with the parallel-tempered \texttt{emcee} code~\cite{Vousden:2016eeu,ForemanMackey:2012ig}. In the present analysis, 750 design points were used. At each design point, one million initial conditions were generated; among these, $40\times 10^3$ events were evolved hydrodynamically; and roughly $200\times 10^3$ \textsc{SMASH} events were produced to maintain sufficient statistics even for 0-1\% collisions (\emph{Trajectum} was initially set up such that these collisions are sampled preferentially \cite{Nijs:2020roc}). 

As explained in \cite{Nijs:2023yab} it is often important to critically assess the full uncertainty matrix. This  includes in particular theoretical uncertainties associated with imperfect modeling, which was in a simplified way taking into account by `observable weighting'. Here we perhaps go for an even simpler option by setting all uncertainties at a minimum value of 4\%. This avoids overfitting the data and we will come back shortly to its effects.

\begin{figure*}[t]
\centering
\includegraphics[width=\textwidth]{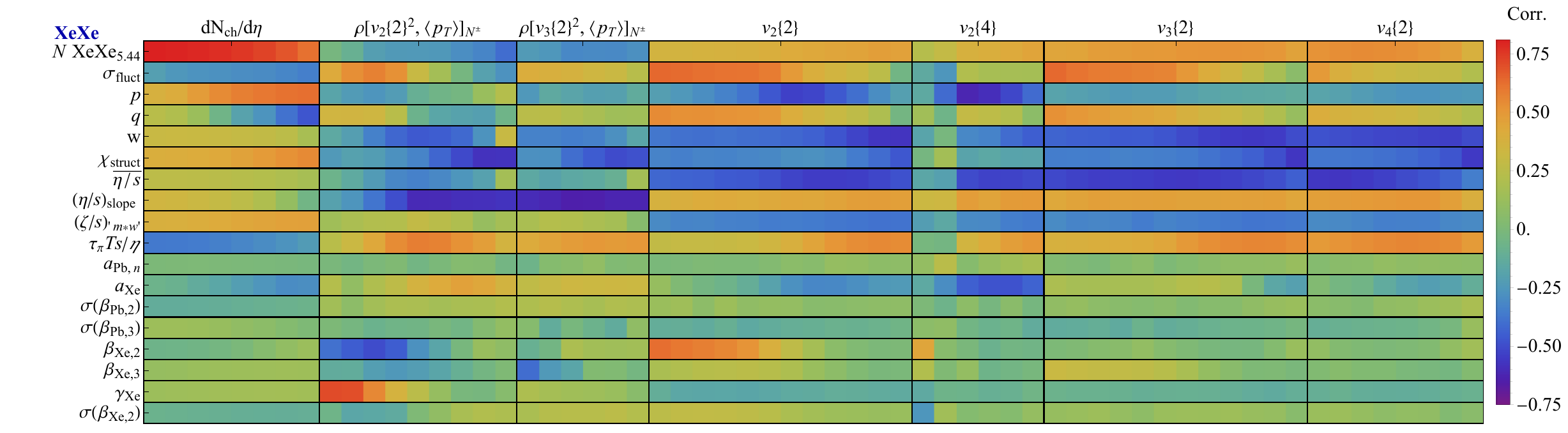}
\includegraphics[width=\textwidth]{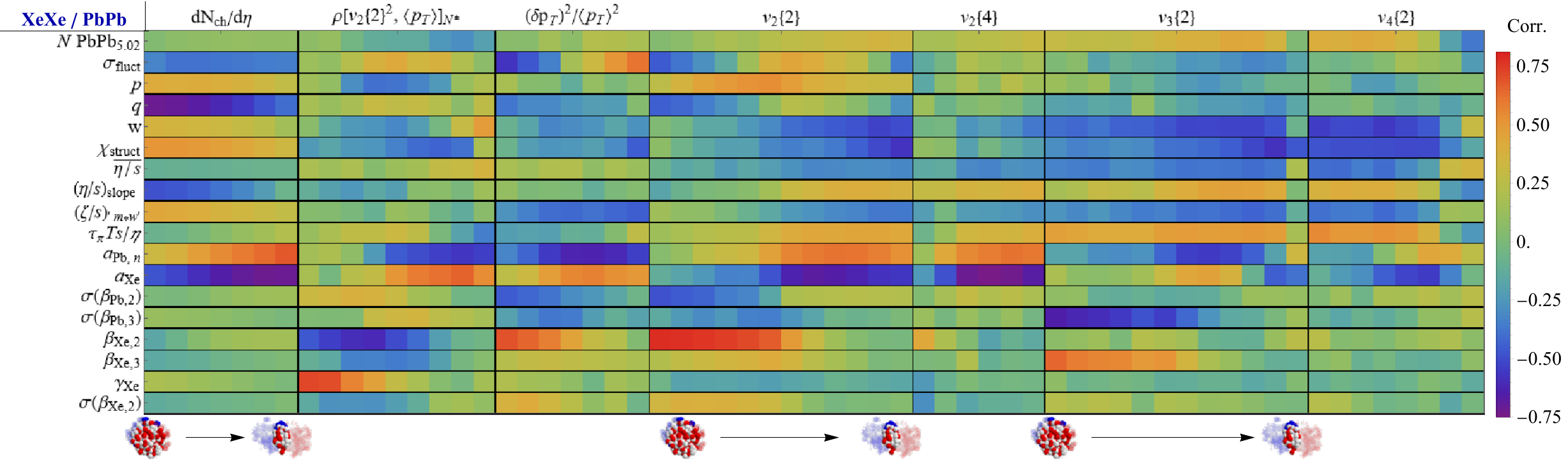}
\caption{\label{fig:designXe-Pb} \textbf{Full sensitivity matrix of observables against model parameters.} We present Xe-Xe collisions (top) and Xe-Xe relative to Pb-Pb collisions (bottom). We display $^{208}$Pb structure parameters in the upper panel to show they do not affect the Xe-Xe observables.
}
\end{figure*}

\paragraph{\textbf{Full correlation matrix and results.}}

In Fig.~\ref{fig:designXe-Pb} we present the same sensitivity matrix shown in Fig.~\ref{fig:2} of the main text, including now the full list of model parameters varied in the analysis, as well as absolute values of observables in Xe-Xe collisions (upper panel). We stress once more the effectiveness of looking at relative quantities to suppress effects related to QGP medium properties. Such a possibility has been argued and documented repeatedly in the literature \cite{Giacalone:2021uhj,Nijs:2021kvn,Xu:2021uar,Zhang:2022fou,Mantysaari:2024uwn,Giacalone:2025vxa}, and is demonstrated here for the first time in a fully systematic calculation.

\begin{figure*}[t]
\centering
\includegraphics[width=\textwidth]{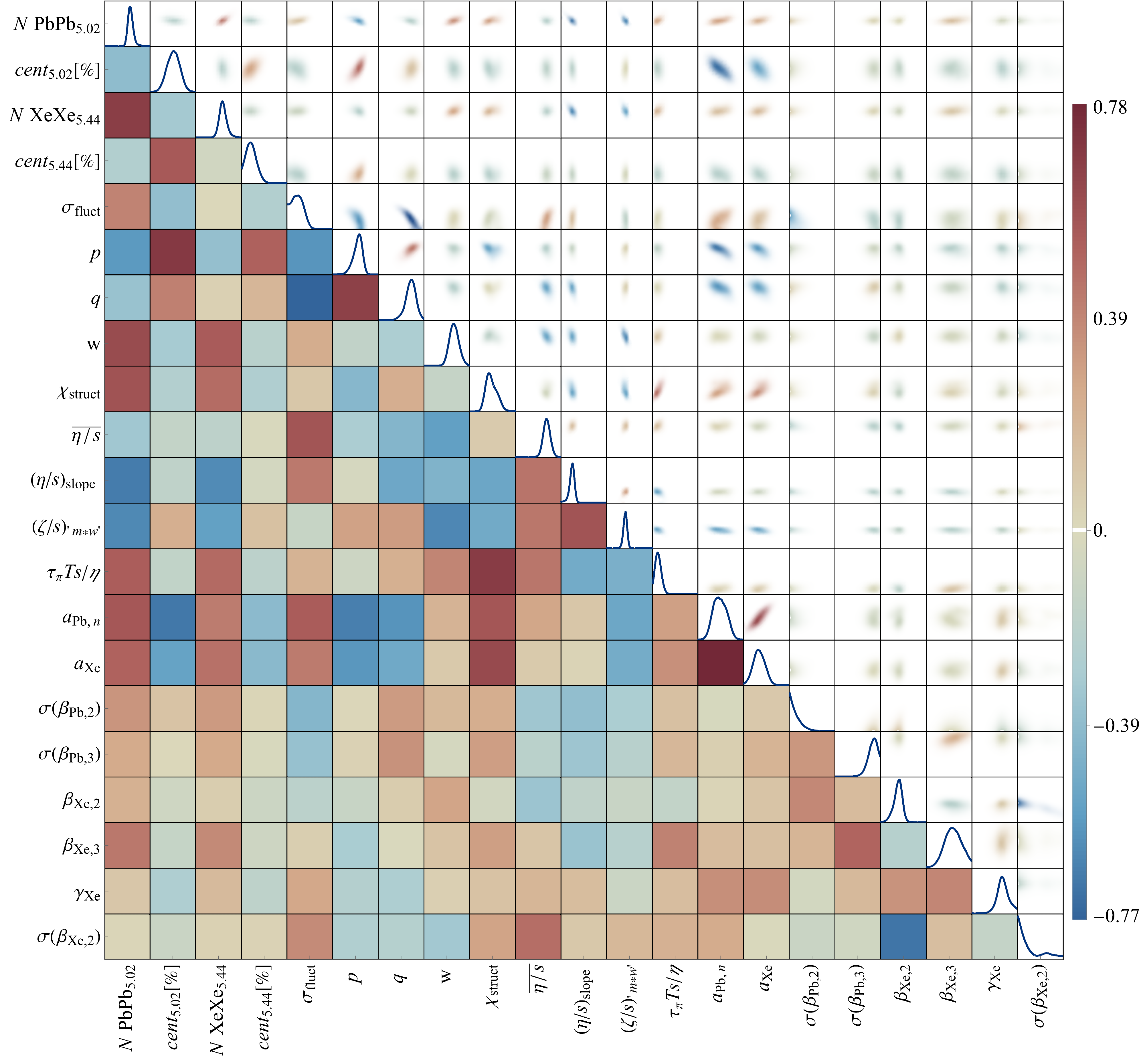}
\caption{\label{fig:fullpost} \textbf{Full posterior correlation matrix in the parameter space.} Prior ranges are provided in Tab.~\ref{tab:params}. 
}
\end{figure*}

In Fig.~\ref{fig:fullpost}, we provide the corner plot displaying the correlations among all model parameters as well as the corresponding posterior distributions. Prior ranges and Maximum a Posteriori (MAP) values for the fitted parameters are listed in Tab.~\ref{tab:params}.

\begin{figure*}[t]
    \centering
    \begin{overpic}[width=\linewidth]{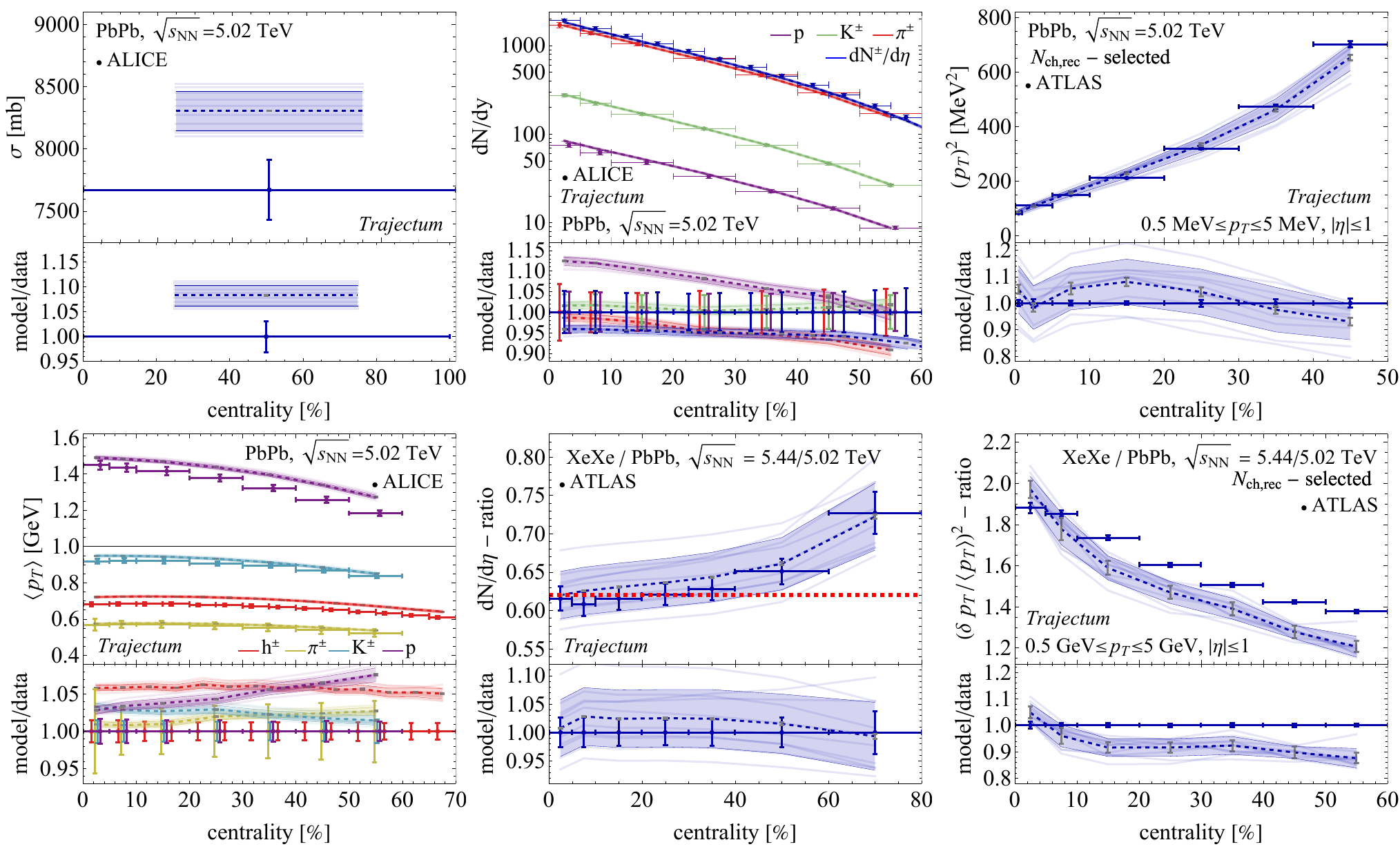}
    \put(30.4,58){\textbf{a}}
    \put(63.5,54){\textbf{b}}
    \put(74,52){\textbf{c}}
    \put(30.4,23.4){\textbf{d}}
    \put(40,24){\textbf{e}}
    \put(97,22){\textbf{f}}
    \end{overpic}\hfill
    \caption{\textbf{Fitted observables.} We show model results for all observables used in the Bayesian fit that were not highlighted in Fig.~\ref{fig:3} of the main text. The red dashed line in panel \textbf{e} corresponds to the ratio of mass numbers, $129/208$, indicating that the multiplicity nicely scales with the number of colliding nucleons. 
}    \label{fig:finalobs}
\end{figure*}

In Fig.~\ref{fig:finalobs} we show comparisons between our model and all other observables that were employed in the Bayesian fit and that have not been shown in the main text.

\paragraph{\textbf{Assessing robustness of results.}}

We want to assess how our results change under two variations of the Bayesian analysis, namely,
\begin{itemize}
    \item by fitting experimental measurements with original uncertainties, rather than inflated to a minimum of 4\% relative errors;
    \item by adding to the fit parameters the half-width radius of the neutron distribution of $^{208}$Pb (while keeping 4\% errors on the experimental results).
\end{itemize}
For completeness, for $^{208}$Pb we split the Woods-Saxon density into a proton density and a neutron density, 
\begin{equation}
\rho^{\rm WS}_{n,p} \propto \left ( 1 + \frac{r-R(R_{n,p};\theta,\phi)}{a_{n,p}} \right )^{-1},
\end{equation}
where the spherical harmonic expansion of $R$ carries the same parameters $\beta_n$ and $\gamma$ for both protons and neutrons. For $^{208}$Pb, our default choices are $a_p=0.448$ fm and $R_p=6.68$ fm, which are constrained by electron scattering experiments at low energy \cite{Loizides:2017ack}. For the neutron half-width radius, our standard choice is $R_n=6.80$ fm, i.e., we implement a shift of 0.12 fm with respect to $R_p$, as suggested by mean-field calculations \cite{Centelles:2010qh}. The value of $a_n$ for $^{208}$Pb is instead a fit parameter. From such an estimate, then, we can evaluate the rms point-proton and point-neutron radii, and estimate the neutron skin $\Delta r_{np}$ of $^{208}$Pb.  For $^{129}$Xe, we implement identical proton and neutron distributions and only attempt at extracting an overall diffuseness parameter $a_{\rm Xe}$, while keeping $R_{\rm Xe}=5.60$ fm, given by mean-field results \cite{Bally:2021qys}. With this setup, we obtain the posterior distributions of $a_{{\rm Pb,}n}$ and $a_{\rm Xe}$ shown in Fig.~\ref{fig:fullpost}.

\begin{figure*}[t]
\centering
\includegraphics[width=0.8\textwidth]{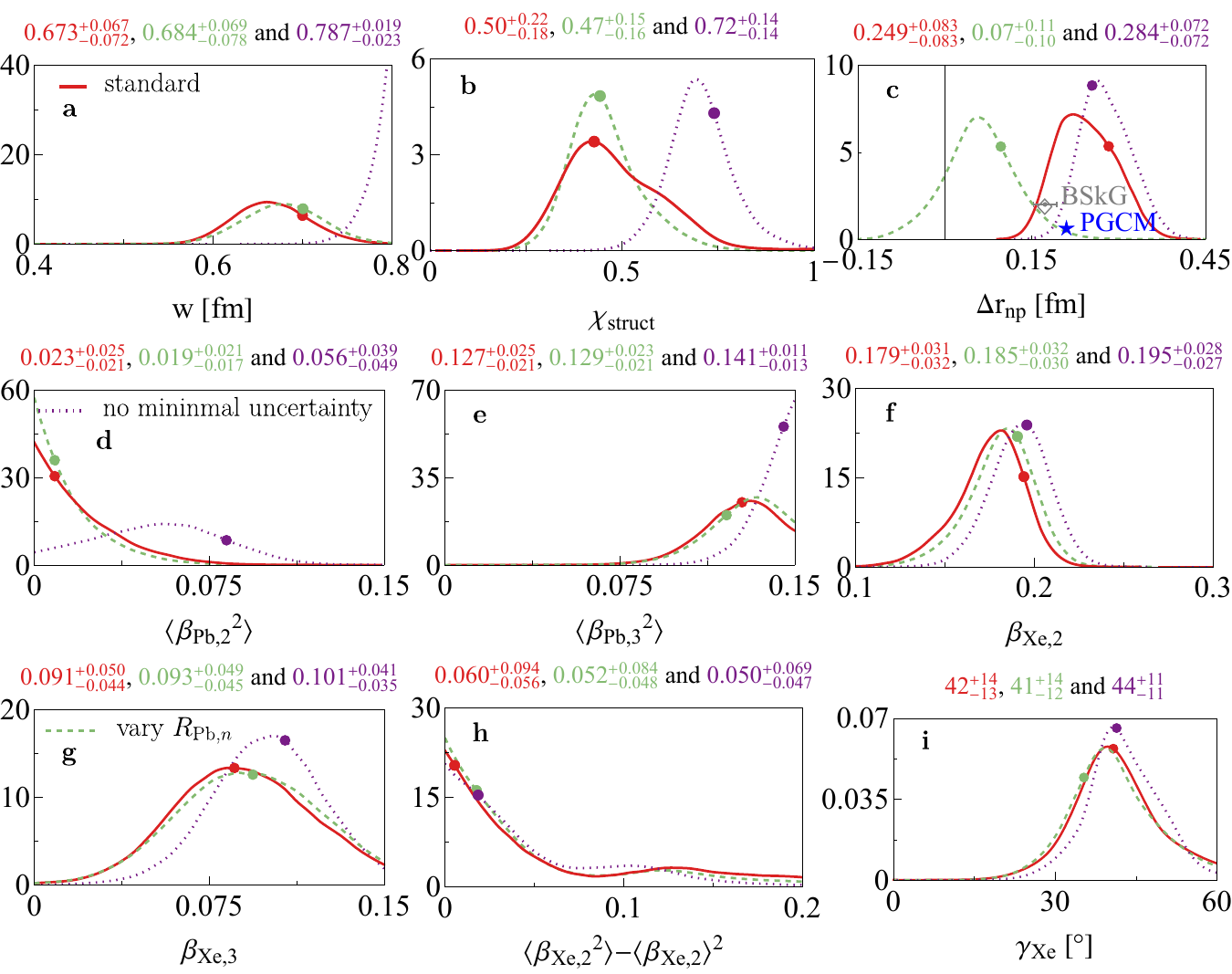}
\caption{\label{fig:varychains}\textbf{Posterior distributions of key initial-state parameters}. They result from three different iterations of the global Bayesian analysis of LHC data. \textbf{a-b}: nucleon structure parameters. \textbf{c}: neutron skin of lead, including the PGCM value of 0.21 fm, and the BSkG1-5 value of $0.173\pm0.020$ fm. \textbf{d-e}: standard deviation of quadrupole and octupole deformation parameters in lead. \textbf{f-i}: axial and triaxial deformations in $^{129}$Xe. The dots represent the MAP values of the inferred parameters (see Tab.~\ref{tab:params}), also reported on top of each panel. Solid lines: Default setup, as discussed in the main text. Dotted line: fitting experimental data with original error. Dashed lines: fitting experimental data with a minimum 4\% relative error and including the half-width radius of $^{208}$Pb, $R_n$, in the fit.
}
\end{figure*}

The results for relevant posterior distributions of parameters are shown in Fig.~\ref{fig:varychains}. Results for our standard setup are shown as red solid lines. 

Considering true experimental uncertainties in the fit, we obtain the posterior distributions depicted as purple dotted lines. Our first observation is that the determination of the deformation parameters introduced for $^{129}$Xe and $^{208}$Pb in Figs.~\ref{fig:varychains}d-i are essentially unchanged compared to the default fit. The determination of the neutron skin of $^{208}$Pb is then in Fig.~\ref{fig:varychains}c. It is consistent with our previous determination \cite{Giacalone:2023cet}, and is also only marginally affected by the lifting of a minimal experimental uncertainty. Dramatic changes are instead observed for the posterior distributions of parameters that model the structure of colliding nucleons, namely, the nucleon size, $w$, and the constituent size, $\chi_{\rm struct}$ (where we set three constituents per nucleon). The nucleon size appears to suffer from the well-documented problem of favoring unreasonably large values in the posterior \cite{Nijs:2022rme,Giacalone:2022hnz}. Remarkably, though, the posterior distribution of the nucleon constituent size, $\chi_{\rm struct}$, exhibits a well-defined peak, substantially improving our previous determination \cite{Giacalone:2023cet}. Likely, this is because our fits now contain the $\rho(v_n\{2\}^2, \langle p_T \rangle)$ observables, that in off-central collisions are sensitive to the granularity of the initial states \cite{Giacalone:2021clp}.

Now, we promote $R_n$ to a fit parameter, while keeping relative experimental errors to a minimum of 4\%. Again, our first remark is that this yields almost no change in the posterior distributions of the deformation parameters in Fig.~\ref{fig:varychains}d-i. Similarly, the nucleon structure parameters are only marginally affected in panels a-b. However, the fit returns an unreasonably small value of the half-width radius, $R_n\approx5.50$ fm, lower than the proton $R_p$ value and inconsistent with the mean-field results. 

Inspecting the sensitivity of observables to $R_n$, this seems to be mostly driven by the total hadronic cross 

section, which naturally depends on the nuclear size, and which seems to be pushing $R_n$ to small values (see e.g. Fig.~\ref{fig:finalobs}).
While this requires further clarification, the crucial consequence is a shift in the neutron skin toward zero. Indeed, with this change the posterior distribution of the skin is essentially consistent with zero in Fig.~\ref{fig:varychains}c. It would be of great help to find observables offering a dedicated sensibility to $R_n$ in $^{208}$Pb. One possibility may come from electroweak boson production in peripheral Pb-Pb collisions, which is sensitive to neutron distributions \cite{Jonas:2021xju}. The rapidity dependence of baryon stopping has also emerged as a potential sensitive probe of the neutron skin \cite{Pihan:2024lxw,Pihan:2025pep,Pihan:2026leq}.

\paragraph{\textbf{Event selection effects on $\rho(v_2\{2\}^2,\langle p_T \rangle)$.}}
In this part, several subtleties in computing the $\rho(v_2\{2\}^2,\langle p_T \rangle)$ correlator are discussed. First, it is important to realize that the correlator depends sensitively on exactly which events are used to evaluate it. This is always done in small centrality classes (standard 1\% in \emph{Trajectum}, typically one or a few particles experimentally). Then it is important how exactly centrality is defined. For the ATLAS analysis, this is typically done using the forward calorimeter (forward $E_T$-selected) or using mid-rapidity tracks (mid-rapidity $N_{\rm ch,rec}$-selected). While the particles used in computing the correlator itself are always corrected for detector effects, this is typically not done when using the centrality selection. 

An advantage of the forward centrality estimator is that there is no self-correlation with the measurement area at mid-rapidity and hence this estimator is often preferred by experiments. For \traj{}, however, this poses a problem, since it is a boost invariant code that is optimized at mid-rapidity. Luckily, the ATLAS collaboration in the auxiliary figures \cite{ATLAS:2022dov}, provided results for both centrality estimators.

\begin{figure*}[t]
    \centering
    \begin{overpic}[width=0.32\linewidth]{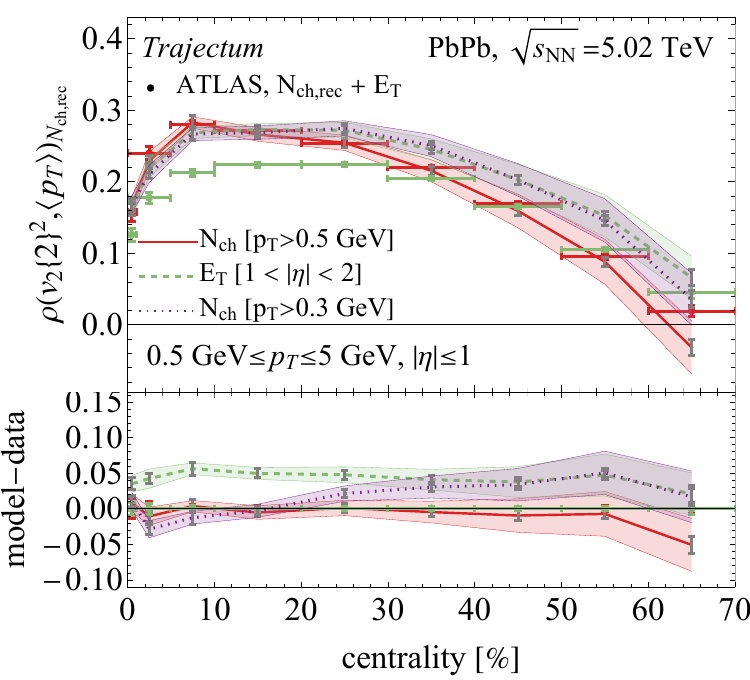}
        \put(92,77){\textbf{a}}
    \end{overpic}\hfill
    \begin{overpic}[width=0.32\linewidth]{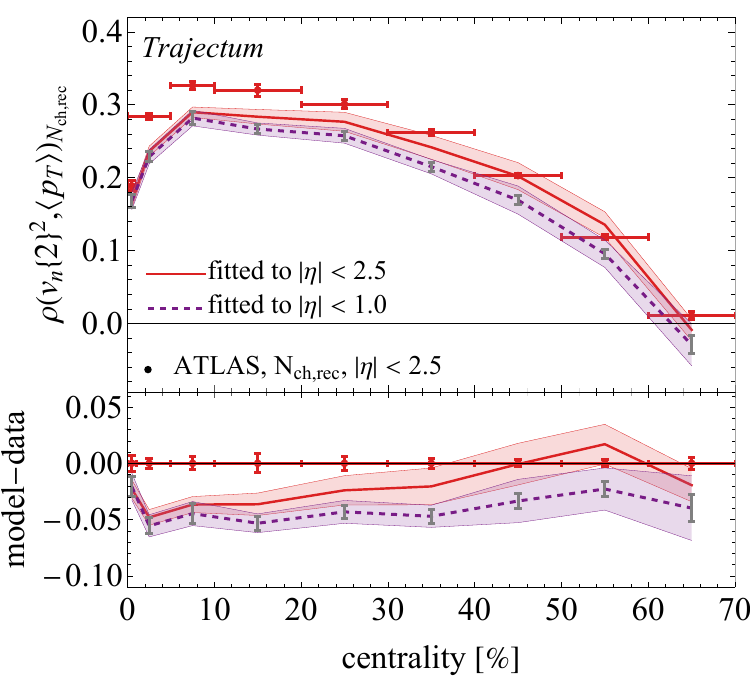}
        \put(92,82){\textbf{b}}
    \end{overpic}\hfill
    \begin{overpic}[width=0.32\linewidth]{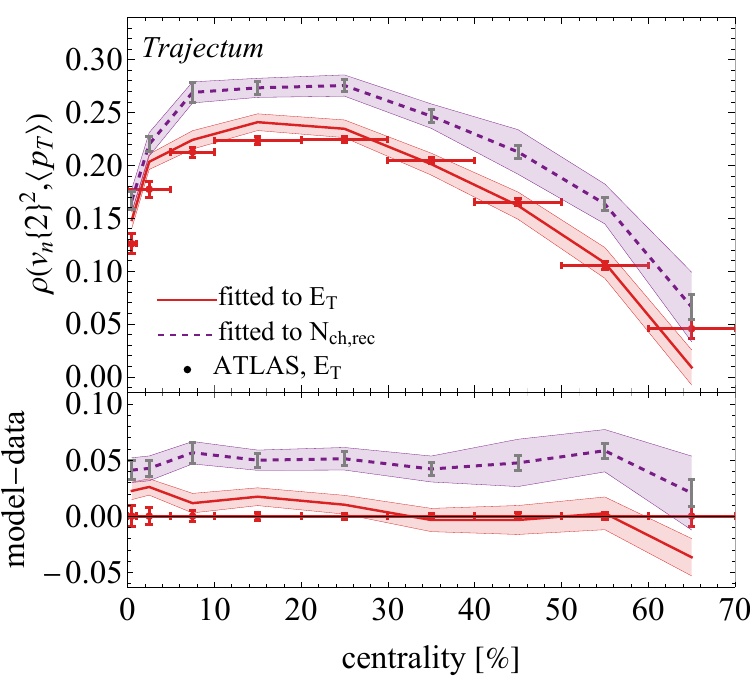}
        \put(92,82){\textbf{c}}
    \end{overpic}
    \caption{\textbf{Systematic bias on the computation of $\rho(v_2\{2\}^2,\langle p_T \rangle)$}. \textbf{a} We show the $\rho(v_2\{2\}^2,\langle p_T \rangle)$ correlator for several centrality selections, both experimentally and from the model. Panels \textbf{b} and \textbf{c} explore a refit of the parameters fitting instead to a larger rapidity range (\textbf{b}) or the forward calorimetric centrality estimator (\textbf{c}). Note that the \emph{Trajectum} evaluation of the observable is modified accordingly, but that we still see that only in (\textbf{c}) a satisfactory fit is possible.
}    \label{fig:allrho2}
\end{figure*}

Fig.~\ref{fig:allrho2} (left) provides ATLAS and \traj{} results for both centrality estimators and for \traj{} an extra mid-rapidity centrality estimator that differs by the lower $p_T$ cut for the charged hadrons (0.3\,GeV as opposed to the standard 0.5\,GeV). Somewhat interestingly, we see that the ATLAS correlator depends strongly on $N_{\rm ch,rec}$-mid-rapidity versus $E_T$-forward for the central bins, while for \traj{} we see a strong difference for peripheral bins (compare the red curves versus the green curves). The combination means that the model misses the data by an almost constant off-set of about 0.05. We also see that for $N_{\rm ch,rec}$-mid-rapidity the lower $p_T$ cut is important for peripheral bins (purple versus red).
\begin{table}[H]
\centering
\caption{\textbf{List of parameter values resulting from the fit of LHC data.} Prior ranges (Lower, Upper) and maximum \emph{a posteriori} (MAP) estimates 
for the model parameters obtained from the Bayesian calibration to Pb--Pb collisions 
at $\sqrt{s_{NN}} = 5.02$~TeV and Xe--Xe collisions at $\sqrt{s_{NN}} = 5.44$~TeV. 
Parameters are grouped by collision-system normalizations, initial-state, nuclear 
structure, and QGP transport properties. The bottom block lists parameters held 
fixed in the present analysis; their values are taken from the MAP estimates of a 
previous Bayesian calibration at $\sqrt{s_{NN}} = 2.76$ and $5.02$~TeV~\cite{Giacalone:2023cet}.}
\label{tab:params}
\small
\renewcommand{\arraystretch}{1.1}
\begin{tabular}{cccc}
\toprule
Parameter & Lower & Upper & MAP \\
\midrule
\midrule
\multicolumn{4}{l}{\textit{Collision-system normalizations}} \\
\midrule
Norm [5.02 TeV] (GeV)              & 14  & 30   & 23.8   \\
$\mathrm{cent}_{\max}$ [5.02 TeV] (\%) & 97  & 103  & 99.3   \\
Norm [5.44 TeV] (GeV)              & 14  & 30   & 23.5   \\
$\mathrm{cent}_{\max}$ [5.44 TeV] (\%) & 97  & 103  & 97.1   \\
\midrule
\multicolumn{4}{l}{\textit{Initial state (T\scriptsize{R}\normalsize{}ENTo)}} \\
\midrule
$\sigma_{\mathrm{fluct}}$          & 0.1  & 1    & 0.399    \\
$p$                                & -0.4 & 0.4  & -0.0498  \\
$q$                                & 1    & 1.45 & 1.28     \\
$w$ (fm)                           & 0.4  & 0.8  & 0.700    \\
$\chi_{\mathrm{struct}}$           & 0    & 1    & 0.428    \\
\midrule
\multicolumn{4}{l}{\textit{Nuclear structure}} \\
\midrule
$a_{\mathrm{Pb},n}$ (fm)           & 0.45 & 0.7  & 0.599    \\
$\sigma(\beta_{\mathrm{Pb},2})$ & 0 & 0.15 & 0.0088 \\
$\sigma(\beta_{\mathrm{Pb},3})$ & 0 & 0.15 & 0.127  \\
$a_{\mathrm{Xe}}$ (fm)           & 0.45 & 0.7  & 0.549    \\
$\beta_{\mathrm{Xe},2}$            & 0.1  & 0.3  & 0.194    \\
$\beta_{\mathrm{Xe},3}$            & 0    & 0.15 & 0.0858   \\
$\gamma_{\mathrm{Xe}}$ (deg)       & 0    & 60   & 40.8     \\
$\sigma(\beta_{\mathrm{Xe},2})$ & 0 & 0.2 & 0.00545 \\
\midrule
\multicolumn{4}{l}{\textit{QGP transport}} \\
\midrule
$(\eta/s)_{\mathrm{ave}}$                          & 0.1 & 0.25 & 0.202    \\
$(\eta/s)_{\mathrm{slope}}$ (GeV$^{-1}$)           & -1  & 2    & -0.261   \\
$(\zeta/s)_{\max\times \mathrm{width}}$ (GeV)      & 0   & 0.02 & 0.00721  \\
$\tau_\pi\, sT/\eta$                               & 1   & 10   & 1.83     \\
\midrule
\multicolumn{4}{l}{\textit{Fixed parameters, mostly taken from \cite{Giacalone:2023cet}}} \\
\midrule
$\sigma_{\rm NN}$ [5.02 TeV] (mb)      & ---  & ---   & 67.6   \\
$\sigma_{\rm NN}$ [5.44 TeV] (mb)      & ---  & ---   & 68.5   \\
$d_{\min}$ (fm)                                    & --- & --- & 0        \\
$R_{\mathrm{Pb},n}$ (fm)                           & --- & --- & 6.80    \\
$n_c$                                              & --- & --- & 2.88     \\
$\tau_{\mathrm{hyd}}$ (fm/$c$)                     & --- & --- & 0.397    \\
$(\eta/s)_{\delta\mathrm{slope}}$ (GeV$^{-1}$)     & --- & --- & -0.393   \\
$(\eta/s)_{0.8\,\mathrm{GeV}}$                     & --- & --- & 0.344    \\
$(\zeta/s)_{\max}$                                 & --- & --- & 0.072    \\
$(\zeta/s)_{T_0}$ (GeV)                            & --- & --- & 0.440    \\
$\tau_{\pi\pi}/\tau_\pi$                           & --- & --- & 2.52     \\
$r_{\mathrm{hyd}}$                                 & --- & --- & 0.798    \\
$T_{\mathrm{switch}}$ (MeV)                        & --- & --- & 154      \\
$a_{\mathrm{EOS}}$                                 & --- & --- & -8.7704    \\
$f_{\mathrm{SMASH}}$                               & --- & --- & 0.953    \\
\bottomrule
\end{tabular}
\end{table}

Figure~\ref{fig:allrho2} (left) uses parameters that are fitted to the $N_{\rm ch,rec}$-mid-rapidity with $|\eta|<1.0$ cut, so it is not surprising that this fits best. It is hence a sensible question if we can fit the $E_T$-forward selection, and also if we can fit while using the (ATLAS preferred) cut of $|\eta|<2.5$. This is explored in Fig.~\ref{fig:allrho2} (middle, right), respectively, for the $\eta$ cut and the $E_T$-forward selection. Perhaps curiously refitting the parameters does not bring the model much closer to the $|\eta|<2.5$ ATLAS results (this data is significantly higher than the $|\eta|<1.0$ cut). Since the $\traj{}$ model is fully calibrated at mid-rapidity (with most data in the range $|\eta|<0.8$) it is not surprising that $\traj{}$ performs better in the narrower range, but it is still interesting that a refit cannot get the correlator to move much closer to the data. Instead, in Fig.~\ref{fig:allrho2} (right) we see that this is in fact possible for the $E_T$-forward selection. Nevertheless, for $\traj{}$ the most realistic apples-to-apples comparison is at the mid-rapidity region since this is where the model works optimally.

\begin{figure}
    \centering
    \includegraphics[width=0.8\linewidth]{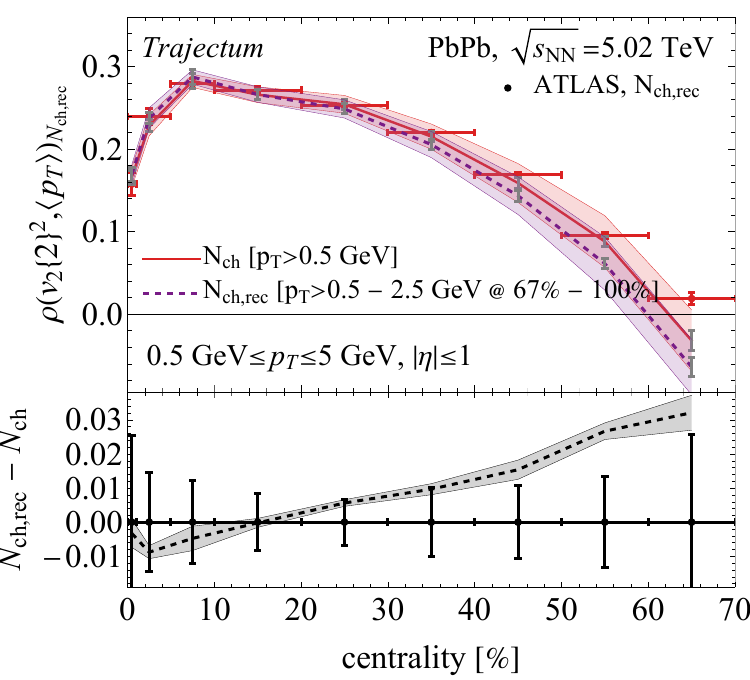}
    \caption{\textbf{Rough estimate of the detector effects present in $\rho(v_2\{2\}^2,\langle p_T \rangle)$.} We assess the effect of using tracks instead of detector-corrected charged hadrons for the mid-rapidity centrality selection. It is a simple estimate (see text for details), but the effect in peripheral bins is of the same order as the systematic uncertainty.
}
    \label{fig:pbrho2detectoreffects}
\end{figure}

Lastly, in Fig.~\ref{fig:pbrho2detectoreffects} we present a simple estimate of the difference in using $N_\text{ch, rec}$ (as ATLAS does) versus fully detector corrected charged hadrons. This is somewhat difficult to estimate, so here we suffice by assuming a 67\% efficiency that linearly increases to 100\% from $0.5 - 2.5\,$GeV. We see that the result is small (barely larger than the \traj{} systematic uncertainty), but note that when comparing with the standard centrality selection in this example we find a difference that is for peripheral bins slightly larger than the ATLAS systematic uncertainty. 

All effects studied in this part are, of course, relatively small. Nevertheless, since we aim to obtain an accurate estimate of the shape of $^{129}$Xe, it is important to ensure that model-to-data comparisons are meaningful, allowing us to report reliable theoretical systematic uncertainties. It is for this reason that in the main text we use the mid-rapidity centrality selector with a narrow rapidity range, which is perhaps unconventional (not the ``preferred'' experimental choice). We stress that the analysis here really benefited from the wide scale of experimental results available from \cite{ATLAS:2022dov}.

\paragraph{\textbf{Evidence of octupole-deformed shapes.}}

Our fits also include octupole deformation parameters, $\beta_3$, for the colliding nuclei. We follow two different implementations for the two nuclei under consideration.
\begin{itemize}
    \item For $^{129}$Xe, we implement a value of $\beta_{\rm Xe ,3}$ that is the same for all nuclei and that we fit from the data.
    \item  For $^{208}$Pb, we instead sample for each simulated nucleus a value of $\beta_3$ from a Gaussian distribution centered at zero and with standard deviation $\sigma ( \beta_{\rm Pb,3}) $, which we also fit from the data.
\end{itemize}
 Quite interestingly, while fluctuations of the quadrupole parameters, $\sigma(\beta_2)$, turn out to be almost negligible for both nuclei, the lead octupole deformation fluctuation is significant and of order $\sigma(\beta_{\rm Pb, 3})\approx 0.1$. The posterior distributions are shown in Fig.~\ref{fig:fullpost}, while MAP values are given in Tab.~\ref{tab:params}. 

This result seemingly validates recent claims that the inclusion of an octupole deformation parameter in the modeling of $^{208}$Pb improves the description of the rms triangular flow coefficient, $v_3\{2\}$, in central collisions \cite{Carzon:2020xwp,Xu:2025cgx}. Quantitative conclusions can be greatly improved in the future by studying the fourth-order cumulant of the triangular flow, $v_3\{4\}$, which is highly sensitive to shape fluctuations \cite{Xu:2025cgx,Liu:2025fnq}. A detailed Bayesian study of the fourth-order cumulant of the $v_3$ distribution is currently beyond our possibilities. Recent advances with neural networks are promising in this respect \cite{Auvinen:2026dyl}.

\begin{table*}[b]
\centering
\begin{tabular}{cccc}
& & $^{208}$Pb & $^{129}$Xe \\
\midrule

\multirow{4}{*}{1-body}
& $\langle R_2\rangle_1\,[\text{fm}^2]$ & $20.92(14)_\text{syst}$ & $15.47(42)_\text{syst}$ \\
& $\langle R_3\rangle_1\,[\text{fm}^3]$ & $115.0(13)_\text{syst}$ & $74.0(33)_\text{syst}$ \\
& $\langle R_4\rangle_1\,[\text{fm}^4]$ & $681(12)_\text{syst}$ & $384(26)_\text{syst}$ \\
& $\langle R_6\rangle_1\,[\text{fm}^6]$ & $28.4(9)_\text{syst}\times10^3$ & $12.6(15)_\text{syst}\times10^3$ \\

\addlinespace[0.5em]
\midrule
\addlinespace[0.5em]

\multirow{2}{*}{2-body}
& $\langle \mathcal{E}_2\,\mathcal{E}_{-2}\rangle_2/\langle R_2\rangle_1^2$ 
& $0.29(3)_\text{stat}(56)_\text{syst}\times10^{-3}$ 
& $13.1(1)_\text{stat}(19)_\text{syst}\times10^{-3}$ \\

& $\langle \mathcal{E}_3\,\mathcal{E}_{-3}\rangle_2/\langle R_3\rangle_1^2$ 
& $-29(38)_\text{stat}\times10^{-6}$ 
& ${4.3(1)_\text{stat}(26)_\text{syst}}\times10^{-3}$ \\

\addlinespace[0.5em]
\midrule
\addlinespace[0.5em]

\multirow{2}{*}{3-body}
& $( ~\langle R_2 \, \mathcal{E}_2\,\mathcal{E}_{-2}\rangle_3~- \langle R_2 \rangle_1\langle \mathcal{E}_2\,\mathcal{E}_{-2} \rangle_2 ~)~ /~
\langle R_2\rangle_1^3$ 
& $1.3(14)_\text{stat}(28)_\text{syst}\times10^{-6}$ 
& $0.01(1)_\text{stat}(30)_\text{syst}\times10^{-3}$ \\

& $ ( ~\langle R_2 \, \mathcal{E}_3\,\mathcal{E}_{-3}\rangle_3~- \langle R_2 \rangle_1\langle \mathcal{E}_3\,\mathcal{E}_{-3} \rangle_2 ~ ) ~ /~
\langle R_2\rangle_1\langle R_3\rangle_1^2$ 
& $-1.4(17)_\text{stat}\times10^{-6}$ 
& $0.139(7)_\text{stat}(75)_\text{syst}\times10^{-3}$ \\

\bottomrule
\end{tabular}
\caption{\label{tab:nuclearproperties}\textbf{One-, two-, and three-body properties of $^{208}$Pb and $^{129}$Xe inferred from the fits of LHC data.} Systematic uncertainties represent the $1\sigma$ variation of these quantities across ten samples of the posterior distributions for the nuclear density parameters. Statistical uncertainties reflect the statistical precision on the hydrodynamic events used in the design of the Bayesian fits.} 
\end{table*}

\paragraph{\textbf{Leading-order estimate of the $v_2\{2\}$ ratio}.}

We now provide a leading-order estimate of the ratio of elliptic flow coefficients between ultra-central \({}^{129}{\rm Xe}+{}^{129}{\rm Xe}\) and
\({}^{208}{\rm Pb}+{}^{208}{\rm Pb}\) collisions that helps clarify how such an observable is linked to the expectation value of the two-body angular correlation of Eq.~(\ref{eq:E2E-2}) of the main text. 

To start, we assume a linear hydrodynamic response between the rms initial ellipticity of the QGP and the final rms elliptic flow,
\begin{equation}
    v_2\{2\} = \kappa_{2}\,\varepsilon_2\{2\} .
\end{equation}
Here $\kappa_{2}$ is the linear hydrodynamic response coefficient, while $\varepsilon_2$ is the second-order eccentricity of the QGP entropy density at the beginning of the hydrodynamic expansion. The previous relation implies, in particular,
\begin{equation}
    \frac{v_2\{2\}_{\rm XeXe}}{v_2\{2\}_{\rm PbPb}} = \frac{\kappa_{2,\rm XeXe}}{\kappa_{2,\rm PbPb}} \frac{\varepsilon_2\{2\}_{\rm XeXe}}{\varepsilon_2\{2\}_{\rm PbPb}}.
\end{equation}
Now, we consider collisions at zero impact parameter in which all nucleons in the colliding ions participate in the interaction. Further, we consider that the total entropy of the QGP is directly proportional to the mass number of the colliding nuclei, which is consistent with both data and comprehensive model-to-data comparisons \cite{Nijs:2023yab}. Under such conditions, for $A\gg1$ and to leading-order in the fluctuation of the entropy density field, the mean-squared elliptic flow reads \cite{Duguet:2025hwi,Bofos:2026huw}
\begin{equation}
    v_2\{2\}^2
    =
     \frac{\kappa_2^2}{2A} \biggl [  \frac{\langle R_4 \rangle_1}{\langle R_2 \rangle_1^2} + A \frac{ \langle \mathcal{E}_2 \, \mathcal{E}_{-2} \rangle_2 }{ \langle R_2 \rangle_1^2 } \biggr],
    \label{eq:eps2_ms}
\end{equation}
where quantum expectation values are taken following the same convention as in Eq.~(\ref{eq:Oavg}) of the main text, with the following normalization for the many-body densities
\begin{equation}
    \int d{\bf r}_1 d{\bf r}_2 \, \rho^{(2)}({\bf r}_1, {\bf r}_2)  = \int d{\bf r} \, \rho^{(1)}({\bf r}) = 1 \,.
\end{equation}
From the hydrodynamic calculations, in the ultra-central collision limit we estimate
\begin{equation}
    \frac{\kappa_{2,\rm XeXe}}{\kappa_{2,\rm PbPb}} \approx 0.95,
\end{equation}
such that, with the values of the nuclear expectation values given in Tab.~\ref{tab:nuclearproperties}, we obtain
\begin{equation}
    \frac{v_2\{2\}_{\rm XeXe}}{v_2\{2\}_{\rm PbPb}} \simeq 1.72. 
\end{equation}
This result aligns nicely with the full 0-1\% Trajectum calculations shown in Fig.~\ref{fig:3} of the main text, and hence even with the experimental results. As expected, the leading-order result lies a little above the full numerical result, as it neglects corrections due to, e.g., nucleon-by-nucleon fluctuations or smearing from the finite nucleon size, which naturally counter the nuclear shape effects. For a large nucleus like $^{129}$Xe, these effects are, however, small, such that we are capable of inferring the normalized two-body expectation $ \langle \mathcal{E}_2 \, \mathcal{E}_{-2} \rangle_2 / \langle R_2 \rangle_1^2$ with a robust and relatively small theoretical error.

\end{document}